\documentstyle[12pt]{article}

\begin{document}
\newcommand{\br}{{\bf r}}
 \title{Maxwell equation, Shroedinger equation, Dirac equation, Einstein
equation defined on the multifractal  sets of the time and the space}
\author{L.Ya.Kobelev   \\
Department of  Physics, Urals State University \\ Lenina Ave., 51 ,
Ekaterinburg 620083, Russia   }
\date{}
\maketitle

 pacs:  01.30.Tt, 05.45, 64.60.A; 00.89.98.02.90.+p. \vspace{1cm}

\section{abstract}
What forms will have an equations of modern physics  if the dimensions of
our time and space are  fractional? The generalized equations enumerated
by title are presented by help the generalized fractional derivatives of
Riemann-Liouville.

\section{Introduction}
In the articles \cite{kob1}-\cite{kob6} the generalized fractional
Riemann-Liouville derivatives (GFD) are determined and the fractal theory
of time and space (and some others physical questions) basing on the using
GFD for functions defined on a multifractal sets are presented. The
multifractal time and space sets are characterized by fractal dimensions
$d_{t}{(\br (t),t)}$  and  $d_{\br}{(t(\bf r),\br)}$. In this paper the
generalization of main equations of the  modern physics are presented for
 the multifractal time and space in the frame of multifractal model of time and
space presented in \cite{kob1}-\cite{kob6}. These equations gives in a
little  corrections for the known equations for the case when the fractal
dimensions (FD) of time $d_{t}$ and space $d_{\br}$ are
$d_{t}=1+\varepsilon(\br (t),t)$ (and so on $d_{\br}$) and the FD are
slightly differs from unity,  i.e. $|\varepsilon|<<1$, that is valid for
small densities of Lagrangians in points $t$, $\br$,  i.e. for weak forces
in the domain of space and time near ${\br, t}$.  All the equations may be
received by means of the principle of minimum of fractal dimensions
functional (see \cite{kob1}) and from this principle the generalized
Euler's equations may be write down. We use more simple method in this
article, consisting in the replacing the ordinary derivatives in the known
equations by GFD (it may be ground by comparison with the generalized
Euler equations). Before receiving the equations we remind the main
definitions and designations of the theory \cite{kob1}-\cite{kob4}:\\
 Generalized fractional derivatives (GFD):\\
We begin from remembering  of the fractional Riemann-Liouville derivatives
definitions \cite{sam1}:
\begin{equation}\label{eq1}
D_{+,t}^{d}f(t)=\left(\frac{d}{dt}\right)^n \int\limits_a^t
{dt'\frac{f(t')}{\Gamma(n-d)(t-t')^{d-n+1}}}
\end{equation}
\begin{equation}\label{eq2} \nonumber
D_{-,t}^{d}f(t)=(-1)^n \left({\frac{d}{dt}} \right)^n\int\limits_t^b
{dt'\frac{f(t')}{\Gamma(n-d) (t'-t)^{d-n+1}}}
\end{equation}

Let a function $f(t)$ of variable $t$ is defined on multifractal set
$S_{t}$   which consist from multifractal subsets $s_{i}({t_{i})}$. We
shall see subsets $s_{i}({t_{i})}$ as the "points" $t_{i}$ (with a
continuous distribution for different multifractal subsets
$s_{i}({t_{i})}$ of multifractal set $S_{t}$ ordered by values of $t$. Let
the function $d(t_{i})=d(t)$ is continuous and describes their fractional
dimensions (in some cases coinciding with local fractal dimensions of set
$S_{t}$ as function $t$. For the elementary generalization the definitions
(\ref{eq1})-(\ref{eq2}) are used physical reasons and variable $t$ is
interpreted as a time. For a continuous functions $f(t)$ (the generalized
functions defined on the class of the finitary functions (see\cite{ge1}),
the fractional derivatives of the Riemann - Liouville are continuous also.
So for infinitesimal intervals of time  the functionals
(\ref{eq1})-(\ref{eq2}) will vary on an infinitesimal quantity. For the
continuous function $d(t)$ the changes it thus also will be infinitesimal.
It allows, as the elementary generalization (\ref{eq1}) that is suitable
for describe the changes the function  $f(t)$ defined on multifractal
subsets $s(t)$ (as well as in the (\ref{eq1})-(\ref{eq2})), to take into
account the summary influence of a kernel of integral
$(t-t')^{-d(t)-n+1}\Gamma^{-1}(n-d(t))$, depending from $d(t)$, on the
$f(t)$ in all points of integration and, instead of
(\ref{eq1})-(\ref{eq2}) to write the integral which takes into account all
this influences. Thus, we enter the following definitions (generalized
fractional derivatives and integrals (GFD)), that account also dependence
$d(t)$ from time and vector parameter $\br(t)$ (i.e. $d_t\equiv
d_t(\br,t)$):
\begin{equation}\label{eq3}
D_{+,t}^{d_t}f(t)=\left(\frac{d}{d_{t}}\right)^n \int\limits_a^t
{dt'\frac{f(t')}{\Gamma(n-d_t(t'))(t-t')^{d_t(t')-n+1}}}
\end{equation}
\begin{eqnarray}\label{eq4} \nonumber
& & D_{-,t}^{d_t}f(t)=(-1)^n \times \\ &\times& \left({\frac{d}{{dt}}}
\right)^n\int\limits_t^b {dt'\frac{{f(t')}}
{{\Gamma(n-d_t(t'))(t'-t)^{d_t(t')-n+1}}}}
\end{eqnarray}
In (\ref{eq3})-(\ref{eq4}), as well as in (\ref{eq1})-(\ref{eq2}), $a$ and
$b$ stationary values defined on an infinite axis (from $-\infty$ to
$\infty$), $a<b$ , $n-1 \leq d_{t}<n$, $n=\{d_{t}\}+1$, $\{d_{t}\}$- the
integer part of $d_{t}\geq 0$, $n=0$ for $d_{t}<0$. The only difference
the (\ref{eq3})-(\ref{eq4}) from the (\ref{eq1})-(\ref{eq2}) is:
$d_{t}=d_{t}(\br (t),t)$- fractional dimensions (further will be used for
it terms " fractal dimensions " (FD) or " the global fractal dimension
(FD)" of subset $s_{t}$) is the function of time and coordinates, instead
of stationary values in the (\ref{eq1})-(\ref{eq2}). Similar to
(\ref{eq1})-(\ref{eq2}), it is possible to define the GFD, (that coincides
for integer values of fractional dimensions $d_{\br}(\br ,t)$ with
derivatives respect to vector variable $\br $) $D^{d_{\br }}_{+,\br }f(\br
,t)$ respect to vector $\br (t)$ variables (spatial coordinates). We pay
attention, that definitions (\ref{eq1})-(\ref{eq2}) are a special case of
Hadamard derivatives \cite{hadam1}.\\

2. The connection between  the fractional dimensions (FD)of time and space
with Lagrangian functions of energy densities read:
\begin{equation}\label{5}
d_{t}=1+\sum_{i,\alpha}\beta_{i,\alpha}L_{i,\alpha}(t,{\mathbf r},
\Phi_{i},\psi_{i})
\end{equation}
In (\ref{5})  $\alpha$ takes value:  $\alpha=t,{\br}$. More complicated
dependencies of $d_{\alpha}$ at $L_{\alpha, i}$ are considered in
\cite{kob1}. Note that relation (\ref{5})  (and similar expression for
$d_{ {\mathbf r} }$ does not contain any limitations on the value of
$\beta_{i}L_{i,\alpha}(t,{\mathbf r},\Phi_{i},\psi_{i})$ unless such
limitations are imposed on the corresponding Lagrangians, and therefore
$d_{t}$ can reach any whatever high or small value.

3. Let's write now the equations enumerated in title in fractal time and
space by  using GFD:

a) Maxwell equations:
\begin{eqnarray}\label{6}
\sum\limits_{i = 1}^3 {} D_{ - ,i,r}^{d_r } D_{ + ,i,r}^{d_r } A_j (x) -
\frac{1} {{c^2 }}D_{ - ,t}^{d_t } D_{ + ,t}^{d_t } A_j (x) + m^2 A_j (x) =
\frac{{4\pi }} {c}j_j (x), \\ j_j  = eD_{ + ,j,t}^{d_j } r_i    \nonumber
\end{eqnarray}
\begin{equation}\label{7}
D_{ + ,j,r}^{d_j } A_j (x) = 0
\end{equation}
In (\ref{6})-(\ref{7}) FD $d_{j}$ is equal to $d_{\br}$ for $j=1,2,3$ and
$d_{t}$ for $j=0$ and introduced the mass of foton for providing existence
of GFD on infinity (then it must be select equal zero).\\\\
 b) Shreodinger equation\\\\
\begin{equation}\label{8}
- i\hbar D_{ + ,t}^{d_t } \psi (\br ,t) =  - \frac{{\hbar ^2 }} {{2m}}D_{
- ,r}^{d_\br } D_{ + ,r}^{d_r } \psi (\br ,t) - e^{2} (\br, t)\psi (\br,t)
\end{equation}
where in $ D_{ - ,r}^{d_r } ,D_{ + ,r}^{d_r }$  operators ${\bf \nabla}$
replaced by operators ${\bf \nabla}  \to {\bf \nabla}  - ie/\hbar c {\bf
A}$)\\\\
 c)Dirac equation\\
\begin{equation}\label{9}
[i\gamma _i (D_{ + ,i}^{d_i }  - ieA_i (x)) - m]\psi (x) = 0
\end{equation}
where $\gamma_{i}$ are Dirac matrices. It is necessary to make the
difference for GFD $D_{i}^{d_{i}}$ with respect to $t$ or with respect to
$\br$ :
\begin{equation}\label{10}
D_i^{d_i }  = D_t^{d_t } ,D_{\br }^{d_r }
\end{equation}
For atomic electrons the main role plays the electric fields of nucleus.
So the density of Lagrangians energy that defined the FG $d_{t}$ may be
selected as
\begin{equation}\label{11}
d_{t}=1+\beta\Phi(\br, t)\approx1+ \frac{e^{2}}{rM_{0}c^{2}}
\end{equation}
where $M_{0}$ is the mass of electrical charge body that originate
electrical field. It is easy to demonstrate that on the distances of the
first Bohr's radius in atoms the fractional corrections to time dimensions
(difference the $d_{t}$ from unity) have values $\sim10^{-8}$, so the
fractal corrections to electron energy $E$ in atoms will be have values
$\sim 10^{-8}E$. It lay out (or in limits domain) of  experimental
possibilities of the modern experiment.\\\\

 d) Einstein equation

It is possible to receive the generalization of general relativity
equation by using two ways. In the first way it is necessary to introduce
a parallel displacement in the Riemann space with fractional dimensions
that may be done without difficulties for weak fields (may be it is
possible to determine the parallel displacement and for strong fields by
the same relations). In that case the carrier of a measure is the Riemann
space and we obtain the determination for covariant derivatives in Riemann
space with fractional dimensions
\begin{equation}\label{12}
D_{ \pm ,\alpha }^{d_{i} } t^{\mu \nu }  = D_{ \pm ,\alpha }^{d_{i} }
t^{\mu \nu }  + \gamma _{\alpha \beta }^{\nu}  t^{\mu \beta } \;\;\;\; i =
t,r
\end{equation}
where $t^{\mu\nu}$ tensor and $\gamma^{\mu\nu}$  metric tensor the Riemann
"four-dimension space with fractional dimensions", $D_{ \pm,\alpha}^{
d_{i}}$ is GFD, $\gamma _{\alpha \beta }^{\nu}$ are Christoffel's symbols
\begin{equation}\label{13}
\gamma _{\alpha \beta }^\nu  = \frac{1} {2}\gamma ^{\nu \sigma}(D_{\pm
,\alpha }^{d_{i}}\gamma _{\beta\sigma } + D_{\pm ,\beta}^{d_{i}}\gamma
_{\alpha \sigma}+D_{\pm ,\sigma }^{d_{i}} \gamma _{\alpha \beta }
\end{equation}
The equations for gravitation field tensor
$\tilde{\Phi}^{\mu\nu}$=${\sqrt{-\gamma}}\cdot{\Phi^{\mu\nu}}$,
${\gamma}=det(\gamma_{\mu\nu})$, $\tilde{t}^{\mu\nu}$=
${\sqrt{-\gamma}}\cdot{t^{\mu\nu}}$, $L$ - is a scalar density of matter
(see in details (\ref{6})-(\ref{7})) than read
\begin{equation}\label{14}
\gamma^{\alpha\beta}D_{-,\alpha}^{\nu,d_{i}}D_{+,\beta}^{\nu,d_{i}}\Phi^{\mu
\nu}+ b^{2}\tilde{\Phi}^{\mu \nu } = - \lambda \frac{\delta L}{\delta
\gamma^{\mu \nu}} = \lambda{\tilde t}^{\mu \nu }
(\gamma^{\mu\nu},\Phi_{A})
\end{equation}
where $b$ is a constant value that necessary to introduce for using more
broad sets of functions with GFD and it after calculations may be put
zero. The equation for curvature tensor (with GFD ) have an usual form
\begin{equation}\label{15}
R^{\mu\nu}-\frac{1}{2}\gamma^{\mu\nu}R=\frac{{8\pi}}{{\sqrt{-g}}}T^{\mu\nu}
\end{equation}
\begin{equation}\label{16}
  D_{\pm,\mu}^{d_i }\tilde g^{\mu \nu} = 0
\end{equation}
The equation (\ref{16}) describes the boundary conditions for $g^{\mu\nu}$
on the Universe surface. Stress, that equations (\ref{12})-(\ref{14})
describe gravitation fields in the Riemann space with fractional
dimensions, i.e. the carrier of measure is the Riemann space. For the case
of weak fields the generalized covariant derivatives may be represented as
(see \cite{kob1})
\begin{equation}\label{17}
D_{\pm,\alpha}^{d_i}t^{\mu\nu}\approx 'D_{\pm,\alpha}^{d_i}t^{\mu\nu} +
''D_{\pm ,\alpha }^{d_i} t^{\mu\nu}
\end{equation}
The $'D_{\pm,\alpha}^{d_i}$ in (\ref{17}) describes the contribution from
FD of time and space, the member  $''D_{\pm,\alpha}^{d_i}$ describes  the
contribution from Riemann space with integer dimensions.

The second way for describing the gravitation fields in the fractal time
and space ( by GFD using ) consists in the other the measure carrier
selection. It is more simplest way to select the measure carrier as the
flat four-dimensions pseudo Euclidean Minkowski space. In that case may be
used (as a base) the system of reference which coincide for FD equal to
unit with Cartesian system of reference (we remember that in the fractal
theory of time and space there are an absolute systems of reference). The
equations of the gravitation in that case will be analogies the equation
of the theory \cite{log1} in which all derivatives replaced on GFD and
metric tensor $\gamma^{\mu\nu}$ are consist the functions (functionals)
originated by fractional dimensions (i.e. it must be the function of $L$ -
Lagrangian energy densities of gravitation fields). These equations have
the form
\begin{equation}\label{18}
\gamma^{\alpha\beta}D_{-,\alpha}^{\nu,d_{i}}D_{+,\beta}^{\nu,d_{i}}\Phi^{\mu
\nu}+ b^{2}\tilde{\Phi}^{\mu \nu } = - \lambda \frac{\delta L}{\delta
\gamma^{\mu \nu}} = \lambda{\tilde t}^{\mu \nu }
(\gamma^{\mu\nu},\Phi_{A})
\end{equation}
The equation (\ref{18}) differs from equation  ( \ref{14}) by three
aspects:

a) the metric tensor $\gamma^{^{\mu\nu}}$ now determined on the Minkowski
space with fractional dimensions;\\ b) it differs  by dependencies of
metrics tensor $\gamma^{\mu\nu}$ from $L$ ( because there are no
dependencies in $\gamma^{\mu\nu}$ of $L$ originating through the Riemann
metric tensor ), there are only dependencies originating through FD;\\ c)
the reason of appearance  of the dependencies the $\gamma^{\mu\nu}$ at $L$
lay in the originate it by the fractal dimensions of time and space. If FG
are integer the (\ref{17}) coincide with equation of the theory
\cite{log1}. For weak fields GFD may be represented only by FD  covariant
derivatives (only one member in right part of (\ref{17})) and in that case
may be represented by metric tensor $g^{\mu\nu}$ of an "effective" Riemann
space with integer dimensions (see \cite{kob1}). We pay attention that the
corresponding results of the theory \cite{log1} for connections between
metric tensor $\gamma^{\mu\nu}$ of Minkowski space with "effective" metric
tensor $g^{\mu\nu}$ of Riemann space and gravitation tensor are the
special case of our theory. In general case the metric tensor of Minkowski
space are complicated function of gravitation field tensor.

We leave for readers  an interesting task to generalize the equations of
quantum gravitation for multifractal time and space.
 \section{Relations between GFD and ordinary derivatives for $d_{\alpha}$ near
 integer values}
 If $d_{\alpha}$$\rightarrow n$ where $n$ is an integer number, for example
   $d_{\alpha}$=$1+\varepsilon(\br (t),t)$, $ \alpha=\br ,t $, in that
 case it is possible represent GFD by approximate relations
 (see \cite{kob1})
\begin{equation}\label{19}
  D_{+,x_{\alpha}}^{1+\epsilon}f(\br (t),t)=
  \frac{\partial}{\partial{x_{\alpha}}} f(\br (t),t)+
  \frac{\partial}{\partial x_{\alpha}}[{\varepsilon
  (\br (t),t)f(\br (t)),t)]}
\end{equation}
The replacement in generalized  Maxwell equations  (\ref{6})-(\ref{7}) ,
 Shroedinger equation (\ref{8}), Dirac equation (\ref{9}), Einstein equation
(\ref{13}) (and so on ) the GFD defined by (\ref{eq3}) by approximation of
GFD defined in the (\ref{19}) gives a possibility to solve numerous tasks
in fractal space and time and calculate the corrections from fractional
dimensions for these tasks.
\section{Conclusion}
1. In this paper were presented the main equation of modern physics
defined on multifractal sets of time and space. In case integer dimensions
of time and space all of them coincide with the  known equation. The value
of correction to integer dimensions of time and space in conditions of
Earth are very small.So, the correction to dimension of time that gives
the gravitational field of Earth on the ground of Earth is equal
$\sim10^{-9}$. The  corrections from electric field nucleus on atomic
distances from nucleus are $\sim 10^{-8}$. So, it may be neglected by
these corrections, but only in the cases of weak fields. In the case of
strong fields all generalized equations becomes in the integral fractional
equations. The last don't have singularities and only this fact originate
the interest to these equations. We pay attention that in the fractal
theory of time and space our Universe is the open system (the statistical
theory of open system are in \cite{klim1},\cite{klim2} \\ 2.Can these
equation be used for strong fields (i.e. for FD that differs a lot of from
integer values). This question now is open. We may only say that answer on
this question concerns with answer on other question: is it possible to
use the method of ordinary Lagrangians of quantum and classical theories
for describing the strong fields? If answer on last question is positive,
there are a hope that the positive answer  for the first question will be
correct.


\begin{thebibliography}{99}
\bibitem{kob6}What dimensions do time and spase have:integer or
fractional? XXX.lanl.gov......
\end{thebibliography}
 \end{document}